\documentclass[aps,amsfonts,prl,twocolumn,showpacs]{revtex4}
\usepackage{epsfig,amsmath,amssymb,bm,epsf,graphics}
\def\Real{{\rm Re\,}}
\def\bra#1{\langle#1\vert}
\def\ket#1{\vert#1\rangle}
\def\ketbra#1{\vert#1\rangle\langle#1\vert}
\def\ipr#1#2{\langle#1\vert#2\rangle}

\def\Longarrow{\protect\@lra}
\def\@lra{\relbar\joinrel\relbar\joinrel\relbar\joinrel%
          \relbar\joinrel\rightarrow}
\def\wstate{{\rm W}}
\def\wtilde{{\widetilde{\rm W}}}

\def\coe#1{{({\rm co}E_{#1})}}
\begin{document}
\title{Geometric measure of entanglement 
for multipartite quantum states}

\author{Tzu-Chieh Wei and Paul M. Goldbart}
\affiliation{Department of Physics, 
University of Illinois at Urbana-Champaign, 
1110 West Green Street, Urbana, Illinois 61801-3080, U.S.A.}

\date{December 4, 2002}
%  \date{\today}

\begin{abstract}
The degree to which a pure quantum state is entangled can be 
characterized by the distance or angle to the nearest unentangled 
state.  This geometric measure of entanglement, already present 
in a number of settings [A. Shimony, Ann. NY. Acad. Sci. 755, p.675 (1995)
and H. Barnum and N. Linden, J. Phys. A: Math. Gen. 34, p.6787 (2001)], 
is explored 
for bipartite and multipartite pure and mixed states.  It is determined for 
arbitrary two-qubit mixed states and for generalized Werner and 
isotropic states, and is also applied to certain multipartite mixed 
states. 
 
\end{abstract}
\maketitle

\noindent
{\it Introduction\/}: 
Only recently, after more than half a century of existence, has the 
notion of entanglement become recognized as central to quantum 
information processings~\cite{NielsenChuang00}.  As a result, the 
task of characterizing and quantifying entanglement has emerged as one 
of the prominent themes of quantum information theory.  There have 
been many achievements in this direction, primarily in the setting of
{\it bipartite\/} systems~\cite{Horodecki01}.  Among these, one highlight 
is Wootters' formula~\cite{Wootters98} for the entanglement of formation 
for two-qubit mixed states; others include corresponding 
results for highly symmetrical states of higher-dimensional 
systems~\cite{VollbrechtWerner01,TerhalVollbrecht00}. 
The issue of entanglement for {\it multipartite\/} states 
poses an even greater challenge, and there have been 
correspondingly fewer achievements: notable examples  
include applications of the 
relative entropy~\cite{VedralPlenio98}, 
negativity~\cite{ZyczkowskiWerner}, and 
Schmidt measure~\cite{EisertBriegel01}.

In this Letter, we present an attempt to face this challenge by 
developing and investigating a certain geometric measure of entanglement, 
first introduced by Shimony~\cite{Shimony95} in the setting of bipartite 
pure states and generalized to the multipartite setting (via projection 
operators of various ranks) by Barnum and Linden~\cite{BarnumLinden01}.  
We begin by examining this geometric measure in pure-state settings, 
and then extend it to mixed states, showing that it satisfies 
certain criteria for good entanglement measures. 
We demonstrate that this measure is no harder to compute than the 
entanglement of formation $E_{\rm F}$, and exemplify this fact  
by giving formulas corresponding to $E_{\rm F}$ for 
(i)~arbitrary two-qubit mixed, 
(ii)~generalized Werner, and 
(iii)~isotropic states. 
We conclude by applying the geometric entanglement measure to certain 
families of multipartite mixed states, for which we provide a practical 
method for computing entanglement, and illustrate this method via two 
examples. 

It is not our aim to cast aspersions on exisiting approaches to 
entanglement; rather we simply wish to add one further aspect to 
the discussion.

\smallskip
\noindent
{\it Basic geometric ideas; application to pure states\/}: We begin 
with an examination of entangled {\it pure\/} states, and how one might 
quantify their entanglement by making use of simple ideas of Hilbert space 
geometry.  Let us start by developing a quite general formulation, appropriate 
for multipartite systems comprising $n$ parts, each of which can have 
a distinct Hilbert space.  Consider a general $n$-partite pure state
$|\psi\rangle=\sum_{p_1\cdots p_n}\chi_{p_1p_2\cdots p_n}
|e_{p_1}^{(1)}e_{p_2}^{(2)}\cdots e_{p_n}^{(n)}\rangle$.
One can envisage a geometric definition of its entanglement 
content via the distance 
$d=\min_{|\phi\rangle}
\Vert\,|\psi\rangle-|\phi\rangle\Vert$
between $\ket{\psi}$ and the nearest separable state $\ket{\phi}$
(or equivalently the angle between them). Here, 
$|\phi\rangle\equiv\otimes_{i=1}^n|\phi^{(i)}\rangle$ 
is an arbitrary separable (i.e., Hartree) $n$-partite pure state, 
the index $i=1\ldots n$ labels the parts, and 
$|\phi^{(i)}\rangle\equiv
\sum_{p_i}c_{p_i}^{(i)}\,|e_{p_i}^{(i)}\rangle$.
Naturally, the more entangled a state is, 
the further away it will be from its best unentangled approximant 
and the wider will be the angle between them. 

To actually find the nearest separable state, it is convenient to 
minimize, instead of $d$, the quantity 
$\Vert|\psi\rangle-|\phi\rangle\Vert^2$, 
subject to the constraint 
$\langle\phi|\phi\rangle=1$. 
In fact, in solving the resulting stationarity condition one may 
restrict one's attention to the subset of solutions $\ket{\phi}$
that obey the further condition that for each factor 
$\ket{\phi^{(i)}}$ 
one has 
$\ipr{\phi^{(i)}}{\phi^{(i)}}=1$.  
Thus, one arrives at the {\it nonlinear eigenproblem\/} for the 
stationary $\ket{\phi}$: 
\begin{subequations}
\label{eqn:Eigen}
\begin{eqnarray}
\!\!\!\!\!\!\!\sum_{p_1\cdots\widehat{p_i}\cdots p_n}
\chi_{p_1p_2\cdots p_n}^*c_{p_1}^{(1)}\cdots\widehat{c_{p_i}^{(i)}}\cdots c_{p_n}^{(n)}=
\Lambda\,{c_{p_i}^{(i)}}^*, \\ 
\!\!\!\!\!\!\!\!\!\!\sum_{p_1\cdots\widehat{p_i}\cdots p_n}\chi_{p_1p_2\cdots p_n} {c_{p_1}^{(1)}}^*\cdots\widehat{{c_{p_i}^{(i)}}^*}\cdots {c_{p_n}^{(n)}}^*=
\Lambda\,c_{p_i}^{(i)}\,,
\end{eqnarray}
\end{subequations} 
where the eigenvalue $\Lambda$ is associated with the Lagrange 
multiplier enforcing the constraint 
$\ipr{\phi}{\phi}\!=\!1$, 
and \,\,$\widehat{}$\,\, denotes exclusion.  
In basis-independent form, Eqs.~(\ref{eqn:Eigen}) read
\begin{eqnarray}
\label{eqn:EigenForm}
\!\!\!\langle\psi|\Big(\mathop{\otimes}_{j(\ne i)}^n|\phi^{(j)}\rangle\Big)
\!=\!\Lambda\bra{\phi^{(i)}}, \ \
\Big(\mathop{\otimes}_{j(\ne i)}^n\langle\phi^{(j)}|\Big)|\psi\rangle
\!=\!\Lambda\ket{\phi^{(i)}}.
\end{eqnarray}
From Eqs.~(\ref{eqn:Eigen}) or (\ref{eqn:EigenForm}) one readily sees 
that the eigenvalues $\Lambda$ are real, in $[-1,1]$, 
and independent of the choice of the local basis 
$\big\{|e_{p_i}^{(i)}\rangle\big\}$. 
Hence, the spectrum  $\Lambda$ can be interpreted as the cosine of the angle 
between $|\psi\rangle$ and $\ket{\phi}$; 
the largest, $\Lambda_{\max}$, which we call the 
{\it entanglement eigenvalue\/}, corresponds to the closest 
separable state.   

Although, in determining the closest separable state, we have used the 
squared distance between the states, there are alternative 
(basis-independent) candidates for entanglement measures: 
the distance, 
the sine, 
or the sine squared of the angle $\theta$ between them 
(with $\cos\theta\equiv\Real\ipr{\psi}{\phi}$). 
We shall adopt $E_{\sin^2}\equiv 1-\Lambda_{\max}^2$ as our
entanglement measure because, as we shall see, when generalizing 
$E_{\sin^2}$ to mixed states we have been able to show that it 
satisfies a set of criteria for entanglement measures.  
We remark that determining the entanglement of $|\psi\rangle$ is 
equivalent to finding the Hartree approximation to the ground-state 
of the auxiliary Hamiltonian
${\cal H}\equiv-|\psi\rangle\langle\psi|$~\cite{ref:Mohit}.

In bipartite applications, the eigenproblem~(\ref{eqn:Eigen}) 
is in fact {\it linear\/}, and solving it is actually equivalent to 
finding the Schmidt decomposition~\cite{Shimony95}.  
Moreover, the entanglement eigenvalue is equal to the square of the maximal 
Schmidt coefficient.  By constrast, for the case of three or more parts, 
the eigenproblem is a {\it nonlinear\/} one, for which one can, in general, 
only address the problem directly, i.e., by determining the eigenvalues and 
eigenvectors simultaneously, presumably numerically.  Yet, as we shall 
illustrate shortly, there do exist certain types of states whose 
entanglement eigenvalues can be determined analytically. 

\smallskip
\noindent
{\it Illustrative examples\/}: 
Suppose we are already in possession of the Schmidt 
decompostion of some two-qubit pure state:  
$|\psi\rangle=
 \sqrt{p}\,|00\rangle
+\sqrt{1-p}\,|11\rangle$.  
Then we can read off the entanglement eigenvalue:  
$\Lambda_{\max}=
\max\{\sqrt{p},\sqrt{1-p}\}$. 
Now, recall~\cite{Wootters98} that the concurrence $C$ for this state 
is $2\sqrt{p(1-p)}$.  Hence, one has
\begin{eqnarray}
\label{eqn:LamConc}
&\Lambda_{\max}^{2}=\frac{1}{2}\left(1+\sqrt{1-C^{2}}\right),
\end{eqnarray}
which holds for arbitrary two-qubit pure states.

The possession of symmetry by a state can alleviate the difficulty 
associated with solving the nonlinear eigenproblem.  To see this, 
consider a state 
$|\psi\rangle=
\sum_{p_1\cdots p_n}\chi_{p_1p_2\cdots p_n}
|e_{p_1}^{(1)}e_{p_2}^{(2)}\cdots e_{p_n}^{(n)}\rangle$  
that obeys the symmetry that the nonzero 
amplitudes $\chi$ are invariant under permutations. 
What we mean by this is that, regardless of the dimensions of the 
factor Hilbert spaces, the amplitudes are only nonzero when the 
indices take on the first $\nu$ values (or can be arranged to do so 
by appropriate relabeling of the basis in each factor) and, moreover, 
that these amplitudes are invariant under permutations of the parties, 
i.e., 
$\chi_{\sigma_1\sigma_2\cdots\sigma_n}=
 \chi_{p_1p_2\cdots p_n}$, 
where the $\sigma$'s are any permutation of the $p$'s. (This symmetry 
may be obscured by arbitrary local unitary transformations.)  
For such states, it seems reasonable to anticipate that the closest 
Hartree approximant retains this permutation symmetry.  Assuming this 
to be the case---and numerical experiments of ours support this 
assumption---in the task of determining the entanglement eigenvalue 
one can start with the Ansatz that the closest 
separable state has the form 
$|\phi\rangle\equiv\otimes_{i=1}^n\big(\sum_j c_j|e_j^{(i)}\rangle\big)$,
i.e., is expressed in terms of copies of a single factor state, 
for which $c^{(i)}_j=c_j$.  To obtain 
the entanglement eigenvalue it is thus only necessary to maximize 
${\rm Re}\,\langle\phi|\psi\rangle$ 
with respect to $\{c_j\}_{j=1}^{\nu}$, a simpler task than maximization 
over the $\sum_{i=1}^{n}d_{i}$ amplitudes of a
generic product state.

To illustrate this, we consider several examples involving 
permutation-invariant states, restricting attention to $\nu=2$. 
The most natural realizations are $n$-qubit systems. 
One can classify these symmetric states, as follows:
\begin{equation}
|S(n,k)\rangle\equiv \sqrt{\frac{k!(n-k)!}{n!}}
\sum_{\rm{\scriptstyle permutations}}
|\underbrace{0\cdots0}_{k}\underbrace{1\cdots1}_{n-k}\rangle.
\end{equation}
As the amplitudes are all positive, one can assume that the closest 
Hartree state is of the form
$|\phi\rangle=\big(\sqrt{p}\,|0\rangle+\sqrt{1-p}\,|1\rangle\big)^{\otimes n}$, 
for which the maximal overlap (w.r.t.~$p$) gives the entanglement
eigenvalue for $|{\rm S}(n,k)\rangle$:
\begin{eqnarray}
\label{eqn:Lambda}
\Lambda_{\max}(n,k)=
\sqrt{\frac{n!}{k!(n\!-\!k)!}}
\left(\frac{k}{n}\right)^{\frac{k}{2}}
{\left(\frac{n-k}{n}\right)}^{\frac{n\!-\!k}{2}}.
\end{eqnarray}
For fixed $n$, the minimum $\Lambda_{\max}$ 
(and hence the maximum entanglement) 
among the $\ket{{\rm S}(n,k)}$'s 
occurs for 
$k=n/2$ (for $n$ even) and 
$k=(n\pm1)/2$ (for $n$ odd). 
In fact, for fixed $n$ the general permutation-invariant state can 
be expressed as
$\sum_k \alpha_k\ket{{\rm S}(n,k)}$ with $\sum_k |\alpha_k|^2=1$.
The entanglement of such states can be addressed via the 
strategy that we have been discussing, i.e.,
via the maximization of a function of (at most) three real parameters.
The simplest example is provided by the $n$GHZ state: 
$|n{\rm GHZ}\rangle\equiv
\big(\ket{{\rm S}(n,0)}+\ket{{\rm S}(n,n)}\big)/\sqrt{2}$.
It is easy to show that (for all $n$)
$\Lambda_{\max}(n{\rm GHZ})=1/\sqrt{2}$ 
and 
$E_{\sin^2}=1/2$.

We now focus our attention on three-qubit settings. Of these, 
the states $\ket{S(3,0)}=\ket{000}$ and $\ket{S(3,3)}=\ket{111}$ 
are not entangled and are, respectively, the components of the 
the 3-GHZ state:
$\ket{{\rm GHZ}}\equiv
\big(\ket{000}+\ket{111})/\sqrt{2}$.
The states 
$\ket{{\rm S}(3,2)}$ 
and 
$\ket{{\rm S}(3,1)}$, 
denoted  
$|{\rm W}\rangle\equiv\ket{{\rm S}(3,2)}=
\big(\ket{001}+\ket{010}+\ket{100}\big)/\sqrt{3}$ 
and 
$\ket{\widetilde{\rm W}}\equiv\ket{{\rm S}(3,1)}=
\big(\ket{110}+\ket{101}+\ket{011}\big)/\sqrt{3}$,
are equally entangled, having 
$\Lambda_{\max}=2/3$ and $E_{\sin^2}=5/9$.  

\begin{figure}[t]
\centerline{\psfig{figure=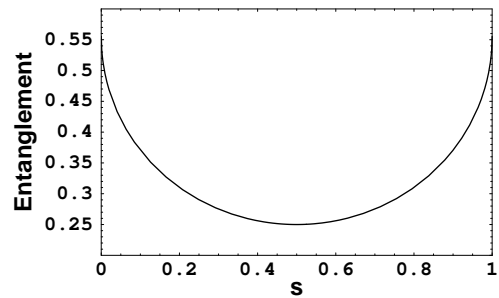,width=6.5cm,height=4.1cm,angle=0}}
\vspace{- 0.4cm}
\caption{Entanglement of the pure state
$\sqrt{s}\,|{\rm W}\rangle+\sqrt{1-s}\,|\widetilde{\rm W}\rangle$ versus $s$. 
This also turns out to be the entanglement curve
for the mixed state $s\,\ketbra{\rm W}+(1-s)\ketbra{\widetilde{\rm W}}$.}
\label{fig:WW}
\vspace{-0.5cm}
\end{figure}
Next, consider a superposition of the ${\rm W}$ and $\widetilde{\rm W}$ 
states: $\ket{\wstate\wtilde(s,\phi)}
\equiv \sqrt{s}\,
\ket{\wstate}+\sqrt{1-s}\,{\rm e}^{i\phi}\ket{\wtilde}$.
It is easy to see that its entanglement is independent of
$\phi$: the transformation 
$\big\{\ket{0},\ket{1}\big\}\to
\big\{\ket{0},{\rm e}^{-i\phi}\ket{1}\big\}$
induces  
$\ket{\wstate\wtilde(s,\phi)}
\rightarrow 
{\rm e}^{-i\phi}\ket{\wstate\wtilde(s,0)}$.
To calculate $\Lambda_{\max}$, assume that the 
separable state is 
$(\cos\theta\ket{0}+\sin\theta\ket{1})^{\otimes 3}$, 
and maximize its overlap with 
$\ket{\wstate\wtilde(s,0)}$.  
Then we find that the tangent $t\equiv\tan\theta$ 
is the particular root of the polynomial 
$\sqrt{1-s}\,t^3+2\sqrt{s}\,t^2-2\sqrt{1-s}\,t-\sqrt{s}=0$
that lies in the range $t\in [\sqrt{{1}/{2}},\sqrt{2}]$.
Via $\theta(s)$, $\Lambda_{\max}$ 
(and $E_{\sin^2}=1\!-\!\Lambda_{\max}^2$) 
can be expressed as 
\begin{eqnarray}
\!\!\!\!&\Lambda_{\max}(s)\!
=\!\frac{1}{2}\big(\sqrt{s}\cos\theta(s)\!
+\!\sqrt{1\!-\!s}\sin\theta(s)\big)
\sin2\theta(s).
\end{eqnarray}
In Fig.~\ref{fig:WW}, we show 
$E_{\sin^{2}}\big(\ket{\wstate\wtilde(s,\phi)}\big)$ vs.~$s$. 
In fact, $\Lambda_{\max}$ of the more general superposition   
\begin{eqnarray}
\label{eqn:SSnk}
\!\!\!\!\!\!\!&\ket{{\rm SS}_{n;k_1k_2}(r,\phi)}\equiv
\sqrt{r}\,                 \ket{{\rm S}(n,k_1)}+
\sqrt{1\!-\!r}\,{\rm e}^{i\phi}\,\ket{{\rm S}(n,k_2)}
\end{eqnarray}
($k_1\ne k_2$) 
turns out to be independent of $\phi$, as in the case of 
$\ket{\wstate\wtilde(s,\phi)}$, and can be computed in the same way.
We note that although the curve in Fig.~\ref{fig:WW} is convex, 
convexity does not hold uniformly over $k_1$ and $k_2$. 

For our last pure-state example, we consider superpositions of 
W and GHZ states: 
$\ket{{\rm WG}(s,\phi)}
\equiv\sqrt{s}\,
\ket{{\rm W}}+\sqrt{1-s}\,\,{\rm e}^{i\phi}\ket{{\rm GHZ}}$.
For these, the phase $\phi$ cannot be ``gauged'' away and,
hence, $E_{\sin^2}$ depends on $\phi$. 
\begin{figure}[t]
\centerline{\psfig{figure=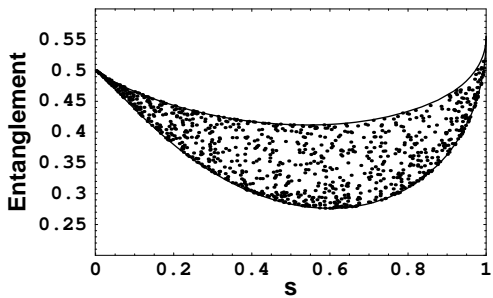,width=6.5cm,height=4.1cm,angle=0}}
\vspace{- 0.4cm}
\caption{Entanglement of 
$\ket{{\rm WG}(s,\phi)}$ versus $s$.  
The upper curve is for $\phi=\pi$ whereas the lower one is for $\phi=0$. 
Dots represent states with randomly generated $s$ and $\phi$.}
\label{fig:WGHZ}
\vspace{-0.5cm}
\end{figure}
In Fig.~\ref{fig:WGHZ} we show $E_{\sin^2}$ vs.~$s$ at $\phi\!=\!0$ and 
$\pi$ (bounding curves), as well as $E_{\sin^2}$ for randomly generated 
values of $s\in[0,1]$ and $\phi\in[0,2\pi]$ (dots).  It is interesting 
to observe that the \lq$\pi$\rq\ state has higher entanglement than 
the \lq$0$\rq\ does.  As the numerical results suggest, the 
($\phi$-parametrized) $E_{\sin^2}$ vs.~$s$ curves of the states 
$\ket{{\rm WG}(s,\phi)}$ lie between the \lq$\pi$\rq\ and \lq$0$\rq\ curves.

\smallskip
\noindent
{\it Extension to mixed states\/}: 
The extension to mixed states $\rho$ can be made via the use of the 
{\it convex roof\/} (or {\it hull\/}) construction [indicated by ``co''], 
as was done for the entanglement of formation 
(see, e.g., Ref.~\cite{Wootters98}).  The essence is a minimization 
over all decompositions $\rho=\sum_i p_i\,|\psi_i\rangle\langle\psi_i|$ 
into pure states, i.e., 
\begin{eqnarray}
\label{eqn:Emixed}
E(\rho)
\equiv
\coe{\rm pure}(\rho)
\equiv
{\min_{\{p_i,\psi_i\}}}
\sum\nolimits_i p_i \, 
E_{\rm pure}(|\psi_i\rangle).
\end{eqnarray}
Now, any good entanglement measure $E$ should, at least, 
satisfy the following criteria 
(c.f.~Refs.~\cite{VedralPlenio98,Horodecki300,Vidal00}):\\
C1.~(a)~$E(\rho)\!\ge\! 0$; 
(b)~$E(\rho)\!=\!0$ if $\rho$ is not entangled.\\
C2.~Local unitary transformations do not change $E$.\\
C3.~Local operations and classical communication (LOCC)
(as well as post-selection, if one wishes) do not increase the 
expectation value of $E$.\\
C4.~Entanglement is convex under the discarding of 
information, i.e., $\sum_i p_i\,E(\rho_i)\ge E(\sum_i p_i\,\rho_i)$.\\ 
The issue of the desirability of additional features, such as 
continuity and additivity, requires further investigation, but 
C1-C4 are regarded as the minimal set, if one is to guarantee 
that one has an {\it entanglement monotone\/}~\cite{Vidal00}.  

Does the geometric measure of entanglement obey C1-4? 
From the definition~(\ref{eqn:Emixed}) it is evident that 
C1 and C2 are satisfied provided that $E_{\rm pure}$ satisfies them, 
as it does for $E_{\rm pure}$ being any function of $\Lambda_{\max}$
consistent with C1.  It is straightforward to check that C4 holds,  
by the convex hull construction.  
The consideration of C3 seems to be more delicate. 
The reason is that our analysis of whether or not it holds 
depends on the explicit form of $E_{\rm pure}$. 
For C3 to hold, it is sufficient to show that the average entanglement is
nonincreasing under any trace-preserving, unilocal operation: 
$\rho\rightarrow \sum_k V_k\rho V_k^\dagger$, where the Kraus
operator has the form  
$V_k=\openone\otimes\cdots \openone\otimes V_k^{(i)}\otimes\openone\cdots\otimes\openone$
and $\sum_k V_k^{\dagger} V_k=\openone$. Furthermore, it suffices to
show that C3 holds for the case of a pure  initial state, i.e., $\rho=\ketbra{\psi}$.  
We now prove that for the particular (and by no means unnatural) 
choice $E_{\rm pure}=E_{\sin^2}$, C3 holds. To be precise,
for any quantum operation 
on a pure initial state, i.e.,~$|\psi\rangle\langle\psi|
\rightarrow\sum\nolimits_k V_k|\psi\rangle\langle\psi|V_k^\dagger$,
we aim to show that
$\sum_k p_k \,E_{\sin^{2}}
\left({V_k|\psi\rangle}/\!{\sqrt{p_k}}\right)
\!\le\!
E_{\sin^{2}}(|\psi\rangle)$,
where 
$p_k\!\equiv\!
{\rm Tr}\,V_k|\psi\rangle\langle\psi|V_k^\dagger
=\langle\psi|V_k^\dagger V_k|\psi\rangle$,
regardless of whether the operation $\{V_k\}$ 
is state-to-state or state-to-ensemble.
Let us respectively denote by $\Lambda$ and $\Lambda_k$ 
the entanglement eigenvalues corresponding to $\ket{\psi}$ 
and the (normalized) pure state ${V_k|\psi\rangle}/{\sqrt{p_k}}\,$. 
Then our task is to show that 
$\sum_k p_k\,\Lambda_k^2 \ge \Lambda^2$,
of which the left hand side is, by the definition of $\Lambda_k$, 
equivalent to 
$\sum_k p_k 
{{\max}\atop{\scriptscriptstyle\xi_k\in D_s}}
\Vert
{\langle\xi_k| V_k|\psi\rangle}/\!{\sqrt{p_k}}
\Vert^2=\sum_k 
{{\max}\atop{\scriptscriptstyle\xi_k\in D_s}}\Vert
\langle\xi_k| V_k|
\psi\rangle\Vert^2$. 
Without loss of generality, we may assume that it is the first party
who performs the operation. Recall that the  
condition~(\ref{eqn:EigenForm}) for the closest separable 
state $\ket{\phi}\equiv|\tilde{\alpha}\rangle_1\otimes|\tilde{\gamma}
\rangle_{2\cdots n}$ can be recast as
${}_{2\cdots n}\langle\tilde{\gamma}|\psi\rangle_{1\cdots n}$$=
$$\Lambda|\tilde{\alpha}\rangle_1$.
Then, by making the specific choice 
$\langle\xi_k|=
({\langle\tilde{\alpha}|V_k^{(1)\dagger}/\!{\sqrt{q_k}}})
\otimes\langle\tilde{\gamma}|$,
where
$q_k
\equiv
\langle\tilde{\alpha}|
V_k^{(1)\dagger}V_k^{(1)}
|\tilde{\alpha}\rangle$,
we have the sought result
$\sum_k p_k\Lambda_k^2=$
$\sum_k {{\max}\atop\xi_k\in D_s}
\Vert\langle\xi_k| V_k|\psi\rangle\Vert^2\ge\Lambda^2\sum_k 
({\langle\tilde{\alpha}|
V_k^{(1)\dagger}V_k^{(1)}|
\tilde{\alpha}\rangle/\!\sqrt{q_k}}
)^2=\Lambda^2$.  
We note that a different approach to establishing this result has been used by 
Barnum and Linden~\cite{BarnumLinden01}.

\smallskip
Before moving on to the {\it terra incognita\/} of mixed {\it multipartite\/}
entanglement, we test the geometric approach 
in the setting of mixed {\it bipartite\/} states, 
by computing $E_{\sin^2}$ for three classes of states for 
which $E_{\rm F}$ is known. 

\smallskip
\noindent
{\it Arbitrary two-qubit mixed states\/}: 
For these we show that 
\begin{eqnarray}
\label{eqn:EC}
& E_{\sin^2}(\rho)=
\frac{1}{2}\big(1-\sqrt{1-C(\rho)^2}\,\big),
\end{eqnarray}
where $C(\rho)$ is the Wootters concurrence of the state $\rho$. 
Recall that in his proof of the formula for 
$E_{\rm F}$, Wootters showed that there exists 
an optimal decomposition $\rho=\sum_{i}p_i\,|\psi_i\rangle\langle\psi_i|$  
in which every $|\psi_i\rangle$ has the concurrence of $\rho$ itself.  
(More explicitly, every $|\psi_i\rangle$ has the identical  
concurrence, that concurrence being the infimum over all  
decompositions.)
By using Eq.~(\ref{eqn:LamConc}) one can, via Eq.~({\ref{eqn:EC}),
relate $E_{\sin^2}$ to $C$ for any two-qubit {\it pure\/} states. 
As $E_{\sin^2}$ is a monotonically increasing function of $C\in[0,1]$, 
the optimal decomposition for $E_{\sin^2}$ is identical to that for 
the entanglement of formation $E_{\rm F}$.  Thus, we see that 
Eq.~(\ref{eqn:EC}) holds for {\it any two-qubit mixed state}.  The fact that $E_{\sin^2}$ is  related to $E_{\rm F}$ via the concurrence $C$ is 
inevitable for two-qubit systems, as both are determined by the one 
independent Schmidt coefficient.  

\begin{figure}[t]
\centerline{\psfig{figure=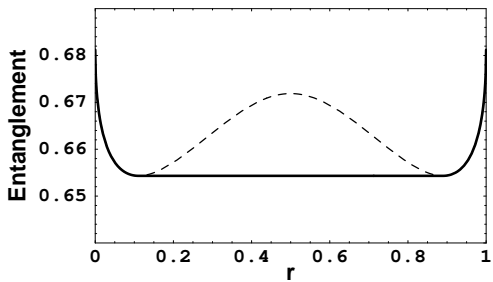,width=6.5cm,height=3.9cm,angle=0}}
\vspace{- 0.4cm}
\caption{Entanglement curve for the mixed state
$\rho_{7;2,5}(r)$ (full line) 
constructed as the convex hull of the curve for the pure state 
$\ket{{\rm S}{\rm S}_{7;2,5}(r,\phi)}$ 
(dashed in the middle; full at the edges).}
\label{fig:SS72}
\vspace{-0.5cm}
\end{figure}
\smallskip
\noindent
{\it Generalized Werner states\/}: 
Any state $\rho_{\rm W}$ of a $C^d\otimes C^d$ system is called a generalized 
Werner state if it is invariant under
${\rm\bf P}_{1}:\rho\rightarrow
\int dU(U\otimes U)\rho\,
(U^\dagger\otimes U^\dagger)$,
where $U$ is any element of the unitary group ${\cal U}(d)$ and $dU$ is 
the corresponding normalized Haar measure.  Such states 
can be expressed as a linear combination of two operators: 
the identity $\hat{\openone}$, and 
the swap $\hat{\rm F}\equiv\sum_{ij}|ij\rangle\langle ji|$, i.e., 
$\rho_{\rm W}\equiv a\hat{\openone}+b\hat{\rm F}$, 
where $a$ and $b$ are real parameters related via the 
constraint 
${\rm Tr}\rho_{\rm W}=1$. 
This one-parameter family of states can be conveniently expressed in 
terms of the single parameter 
$f\equiv{\rm Tr}(\rho_{\rm W}\hat{\rm F})$.
By employing the technique by developed by Vollbrecht 
and Werner~\cite{VollbrechtWerner01} 
[as applied to $E_{\rm F}(\rho_{\rm W})$]
to $E_{\sin^2}$, 
one arrives at
the geometric entanglement function for Werner states: 
\begin{eqnarray}
 \label{eqn:Werner}
 & E_{\sin^2}\big(\rho_{\rm W}(f)\big)=
 \frac{1}{2}\big({1-\sqrt{1-f^2}}\,\big)\quad {\rm for} \ f\le 0,
 \end{eqnarray}
 and zero otherwise.

\smallskip
\noindent
{\it Isotropic states\/}: 
These are states invariant under 
${\rm \bf P}_{2}:\rho\rightarrow 
\int dU\,(U\otimes U^*)\rho\,
(U^\dagger\otimes {U^*}^\dagger)$, and can be expressed as 
\begin{eqnarray}
\rho_{\rm iso}(F)
\equiv
\frac{1-F}{d^2-1}
\left(\hat{\openone}-|\Phi^+\rangle\langle\Phi^+|\right)+
F|\Phi^+\rangle\langle\Phi^+|,
\end{eqnarray}
where $|\Phi^+\rangle\equiv\frac{1}{\sqrt{d}}\sum_{i=1}^{d}|ii\rangle$ and
$F\in[0,1]$. For $F\in[0,1/d]$, this state is known to be separable~\cite{Horodecki299}.
By using the Vollbrecht-Werner technique and following 
arguments similar to those of
Terhal and Vollbrecht~\cite{TerhalVollbrecht00}
applied to $E_{\rm F}(\rho_{\rm iso})$, one arrives 
(for $F\ge 1/d$) at 
\begin{eqnarray}
& \!\!\!\!\!\! E_{\sin^2}\left(\rho_{\rm iso}(F)\right)=
1-\frac{1}{d}\big(\sqrt{{F}}
+\sqrt{(\!1-\!F)(d\!-\!1)}\,\big)^2.
\end{eqnarray}
\noindent
{\it Mixtures of multipartite symmetric states\/}: 
As a final example we consider mixed states of the form ($k_1\ne k_2$)
\[\rho_{n;k_1\!k_2}\!(r)\!\equiv\! r\, 
\ket{{\rm S}(n,k_1)\!}\bra{{\rm S}(n,k_1)\!}+
(1-r)\ket{{\rm S}(n,k_2)\!}\bra{{\rm S}(n,k_2)\!}.\] 
From the independence of 
$E_{\sin^2}\left(\ket{{\rm S}{\rm S}_{n;k_1k_2}(r,\phi)}\right)$ 
on $\phi$, one can show that 
$E_{\sin^2}\left(\rho_{n;k_1k_2}(r)\right)$ vs.~$r$ can be 
constructed from the convex hull of the entanglement function of 
$\ket{{\rm S}{\rm S}_{n;k_1k_2}(r,0)}$ vs.~$r$.  An example is 
shown in Fig.~\ref{fig:SS72}. 
If the dependence of $E_{\sin^2}$ on $r$ is already convex for the 
pure state, its mixed-state counterpart has precisely the same dependence.  
Figure~\ref{fig:WW}, for which $(n,k_1,k_2)=(3,1,2)$, 
exemplifies such behavior. 
  
\smallskip
\noindent
{\it Concluding remarks\/}: 
We have considered a general, geometrically motivated measure 
of entanglement, applicable to pure and mixed quantum states involving 
arbitrary numbers and structures of parties.  We have illustrated this 
measure via several examples.  In bipartite settings, this approach 
provides an---in general, inequivalent---alternative to the entanglement 
of formation; it is, moreover, naturally extendable to multipartite 
settings.

%\smallskip
\noindent
{\it Acknowledgments\/}: 
We thank J.~Eisert, P.~Kwiat, D.~Leung, S.~Mukhopadhyay, 
M.~Randeria and especially W.~J.~Munro for discussions.
PMG acknowledges the hospitality of the 
Unversity of Colorado--Boulder and the Aspen Center for Physics. 
This work was supported by 
NSF EIA01-21568 and 
DOE DEFG02-91ER45439.
TCW acknowledges a Mavis Memorial Fund Scholarship. 

\end{document}